\documentclass[aps,twocolumn,prl,showpacs]{revtex4}
\usepackage{graphicx,amssymb,amsmath}
\usepackage{graphicx}
\usepackage{dcolumn}
\usepackage{bm}

\newcommand{\bpar}{$B_{||}$}
\newcommand{\bperp}{$B_{\perp}$}

\begin{document}
\title{Quantum Hall Spin Diode}
\author{J. P. Eisenstein}
\affiliation{Institute for Quantum Information and Matter, Department of Physics, California Institute of Technology, Pasadena, CA 91125, USA}
\author{L. N. Pfeiffer}
\affiliation{Department of Electrical Engineering, Princeton University, Princeton, NJ 08544, USA}
\author{K. W. West}
\affiliation{Department of Electrical Engineering, Princeton University, Princeton, NJ 08544, USA}

\date{\today}

\begin{abstract}
Double layer two-dimensional electron systems at high perpendicular magnetic field are used to realize magnetic tunnel junctions in which the electrons at the Fermi level in the two layers have either parallel or anti-parallel spin magnetizations.  In the anti-parallel case the tunnel junction, at low temperatures, behaves as a nearly ideal spin diode.  At elevated temperatures the diode character degrades as long-wavelength spin waves are thermally excited.  These tunnel junctions provide a demonstration that the spin polarization of the electrons in the $N=1$ Landau level at filling factors $\nu = 5/2$ and 7/2 is essentially complete, and, with the aid of an in-plane magnetic field component, that Landau level mixing at these filling factors is weak in the samples studied.

\end{abstract}
\maketitle

Beginning with the pioneering experiments of Tedrow and Meservey \cite{tedrow71,meservey94}, electron spin-dependent tunneling has grown into an enormous field of both technological and fundamental scientific importance.   For example, today tunnel junctions consisting of two ferromagnetic electrodes separated by a thin insulating barrier are the basic element in magnetic random access memories, and these spintronic devices have become increasingly competitive with conventional memories.  On the fundamental physics side, Tedrow and Meservey \cite{tedrow71} made elegant use of their observed spin Zeeman splitting of the superconducting energy gap in thin aluminum films \cite{tedrow70} to explore spin polarized tunneling currents in junctions having Al and ferromagnetic nickel electrodes.  These early experiments opened a new door to the study of the spin structure of magnetic materials. 

Here we explore spin-dependent tunneling in magnetic tunnel junctions fabricated from bilayer two-dimensional electron systems (2DES) in GaAs/AlGaAs heterostructures.  In the presence of a large magnetic field \bperp\ perpendicular to the 2D planes, the energy spectrum of a clean and non-interacting 2DES consists of a ladder of discrete Landau levels (LLs), each of which is split into two spin sublevels by the Zeeman effect.  In a bilayer 2DES, proper adjustment of the magnetic field (which sets the degeneracy $eB_{\perp}/h$ of the spin-split LLs) and the individual 2DES densities $n_1$ and $n_2$ can realize a magnetic tunnel junction in which the spin polarization of the electrons at the Fermi level in the two 2DES layers are either parallel or antiparallel.   For example, if the Landau level filling fraction $\nu=nh/eB_{\perp}$ in both layers is set to $\nu_1=\nu_2=1/2$, the two 2DESs would be fully spin polarized and have parallel magnetizations.  Alternatively, if the densities are adjusted so that $\nu_1$=1/2 but $\nu_2=3/2$, then the electron spins at the Fermi level in the two layers are oppositely directed.  Absent spin-flip tunneling processes, the tunneling conductance at zero bias would be singularly \cite{singular} large in the 1/2-1/2 case but zero in the 1/2-3/2 set-up.  Only at an interlayer voltage equal to the Zeeman splitting would tunneling occur in the latter case.

In real 2DESs at high magnetic fields the above scenario is enriched dramatically by electron-electron interactions.  Instead of sharp spin-split Landau levels, the spectral functions of the 2DES are heavily broadened by interactions.  This broadening is directly observable in the current-voltage ($IV$) characteristic for tunneling between two parallel 2DESs \cite{eisenstein92}.  In addition, these same interactions suppress the tunneling conductance $dI/dV$ around zero bias, creating a Coulomb pseudogap at the Fermi level.  The pseudogap arises from the inability of a correlated 2DES at high magnetic field to rapidly relax the charge defect created by the injection (or extraction) of a tunneled electron \cite{ashoori90,eisenstein92,hatsugai93,he93,johannson93,efros93,varma94,haussmann96,levitov97}.  

Electron-electron interactions can also profoundly affect the spin configuration of a 2DES, allowing it to deviate qualitatively from that expected via simple Pauli counting rules \cite{halperin83,perspectives}.  This is not surprising since in a GaAs-based 2DES the spin Zeeman energy is typically of order $E_Z\sim1$ K while the mean Coulomb energy $E_C$ between electrons in a partially filled LL is typically 10 to 100 times larger \cite{energies}.  In perhaps the most dramatic example, the spin polarization of the 2DES at $\nu = 1/2$ in the lowest Landau level is known to be incomplete at low density \cite{kukushkin99,dementyev99,melinte00,dujovne05,tracy07,li09}.  

In this paper we report tunneling measurements on density imbalanced bilayer 2D electron systems, focussing on the case where $\nu_1=5/2$, while $\nu_2=7/2$; both filling factors being in the first excited, $N=1$ LL.  Our measurements reveal that at low temperature the tunneling $IV$ characteristic is extremely asymmetric, behaving as a nearly ideal diode at low voltage.  This demonstrates that the spin polarization of the electrons in the partially filled $N=1$ LL is nearly complete in both 2D layers.  Elevated temperatures rapidly degrade this diode-like behavior, with thermal excitation of long-wavelength spin waves being likely responsible. 
 
\begin{figure}
\includegraphics[width=3.25in]{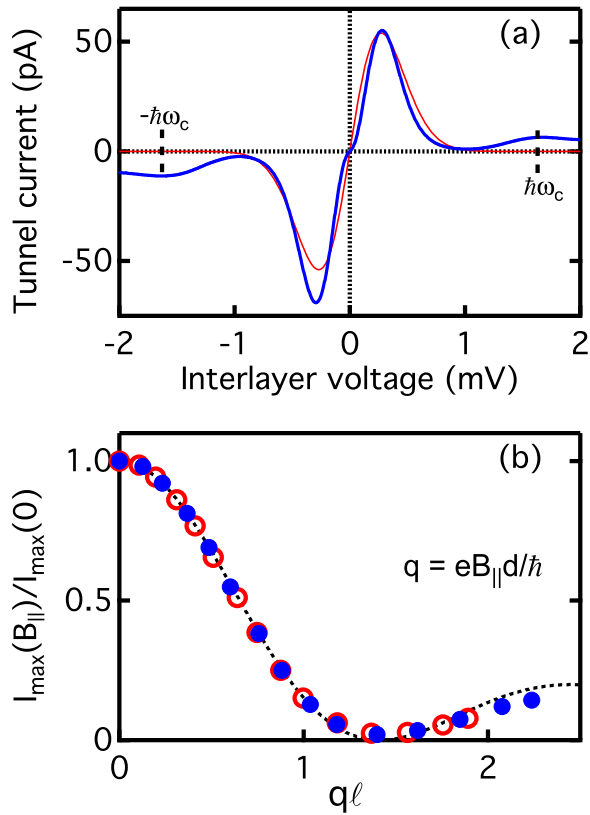}
\caption{\label{}(color online) (a) Blue: Tunneling $IV$ characteristic at $B_{\perp}=0.94$ T and $T=50$ mK with both 2DESs at $\nu = 5/2$. Light red: Representative theoretical $IV$ curve for simple gaussian spectral functions at zero temperature. (b) Observed effect of an in-plane magnetic field $B_{||}$ on the normalized maximum tunneling current at $\nu=5/2$ (solid blue circles, \bperp\ = 0.94 T) and $\nu = 7/2$ (open red circles, \bperp\ = 0.67 T).  Dashed curve: Matrix element effect described in text.}
\end{figure}
The samples and methods we employ for measuring tunneling between parallel 2DESs have been described in detail previously \cite{eisenstein90,eisenstein91}.  In capsule, a double layer 2DES is created by molecular beam epitaxial growth of a GaAs/AlGaAs double quantum well heterostructure. Each 18 nm wide GaAs quantum well contains a 2DES of nominal density $5\times 10^{10}$ cm$^{-2}$ and low temperature mobility $10^6$ cm$^2$/Vs.  These wells are separated by a 10 nm thick Al$_{0.9}$Ga$_{0.1}$As tunnel barrier, leaving the center-to-center separation of the quantum wells at $d=28$ nm. Independent control over the electron density in each layer is enabled by electrostatic gates on the top and backside of the sample. The 2DESs are confined to a 250$\times$250 $\mu$m square region, with arms extending to ohmic contacts to the individual 2D layers.  These contacts enable direct measurements of the tunneling current $I$ flowing in response to an applied interlayer voltage $V$. 

Figure 1(a) shows a typical high magnetic field tunneling $IV$ characteristic for the situation where the two 2DES layers have the same electron density.  For these data, taken at \bperp\ = 0.94 T and $T = 50$ mK, the Landau level filling factor in each 2DES is $\nu = 5/2$.  (The present samples are of insufficient quality for the $\nu = 5/2$ fractional quantized Hall effect to be observed.)  The data in Fig. 1(a) exhibit the main consequences of strong electron-electron interactions at high magnetic field discussed above: a broad peak and valley in the tunneling current, roughly \cite{asymmetry} anti-symmetric in voltage, and a suppression of the tunneling conductance $dI/dV$ very close zero bias.  The width of the main tunneling peaks, $\Gamma \approx 0.3$ mV, is comparable to the mean Coulomb repulsion, $E_{C,v}=e^2 n_v^{1/2} /\epsilon \approx 1$ meV, between the $n_v=eB_{\perp}/2h$ electrons (or holes) in the partially occupied $N=1$ LL.  The suppression of $dI/dV$ close to $V=0$ is the Coulomb pseudogap mentioned above.  Weak at this low magnetic field, this effect is not part of our present focus. The thin red trace in the Fig. 1(a) shows an $IV$ curve calculated assuming simple gaussian spectral functions for the valence $N=1$ LL in each 2DES.

Before turning to the effects of electron spin on the tunneling $IV$ characteristics, we explore the {\it orbital} character of the energy levels involved.  At the magnetic fields studied here the cyclotron splitting $\hbar \omega_c=\hbar e B_{\perp}/m^*$ between Landau levels is not much larger than the interaction-induced broadening of those levels.  At $\nu = 5/2$, for example, the partially filled valence energy level may be dominated by the $N=1$ LL but have other LLs mixed in by interactions.  This possibility is highlighted by the presence, in Fig. 1(a), of weak additional extrema in the tunnel current around $|V| = \pm 1.6$ mV, close to the cyclotron energy $\hbar \omega_c =1.63$ meV at this magnetic field.  Such inter-LL tunneling events are forbidden \cite{forbidden} in a clean, non-interacting 2DES \cite{disorder}.

To address this question, an in-plane magnetic field \bpar\ is added to the perpendicular field \bperp.  Landau level  tunneling matrix elements evolve in a systematic way with the momentum boost $q=eB_{||} d/\hbar$ created by the in-plane field \cite{hu92}.  For tunneling between states solely within the $N=1$ Landau level, this effect, taken alone, requires the tunnel current to depend on \bpar\ as $I(B_{||})=I(0) (1-q^2\ell^2/2)^2 e^{-q^2\ell^2/2}$, where $I(0)$ is the tunnel current at \bpar=0 and $\ell=(\hbar/eB_{\perp})^{1/2}$ is the magnetic length.  Interestingly, the factor $(1-q^2\ell^2/2)^2$ forces the tunnel current to vanish at $q\ell=\sqrt{2}$.  This vanishing is a direct result of the node in the $N=1$ Landau level wavefunction \cite{eisenstein16}.

\begin{figure}
\includegraphics[width=3.25in]{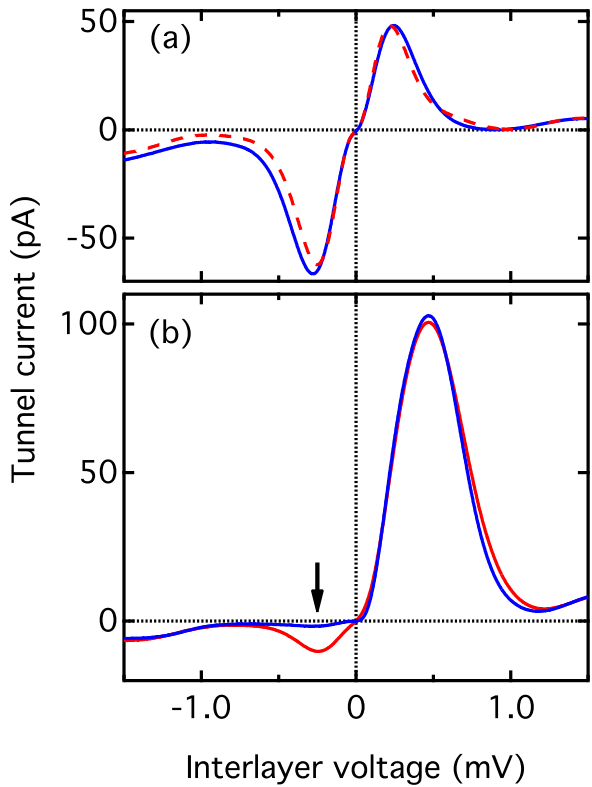}
\caption{\label{}(color online) (a) Density balanced tunneling $IV$ curves at $\nu=5/2$ (blue) and $\nu = 7/2$ (dashed red) at \bperp\ = 0.84 T and $T=50$ mK. (b) Diode-like tunneling response observed at \bperp\ = 0.84 T with one 2D layer at $\nu = 5/2$ and the other at $\nu = 7/2$.  $T=50$ mK (blue) and 200 mK (red).}
\end{figure}

Figure 1(b) shows the dependence of the maximum tunnel current \cite{maxdef} at $\nu=5/2$ and 7/2 on $q$ (and therefore \bpar), normalized by its value at \bpar=0.  The close agreement with the matrix element effect just described (dashed line in the figure) demonstrates that the tunnel current at voltages $|V|\lesssim 1$ mV is dominated by hopping events between states solely within the $N=1$ Landau level.  Landau level mixing, whether due to interactions or disorder, is apparently weak.

The results discussed thus far offer little insight into the spin structure of the partially occupied $N=1$ Landau level.  To expose this structure we turn to tunneling measurements in which the layer densities are adjusted to simultaneously produce $\nu=5/2$ in one layer and $\nu = 7/2$ in the other.  If the spin configuration of the 2DES follows simple Pauli counting rules then the $\nu = 5/2$ layer has only `up' spins at the Fermi level while the $\nu = 7/2$ layer has only `down' spins.  In the 5/2 layer the `down' spin branch is completely empty while in the 7/2 layer the `up' spin branch is completely full.  If electrons tunneling from one layer to the other preserve their spin, the tunneling $IV$ curve should be highly asymmetric. At zero temperature no electrons can tunnel from the 5/2 layer into the 7/2 layer since all up-spin final states are occupied.  In contrast, tunneling from the 7/2 layer to the 5/2 layer can readily proceed as there are empty states in both spin bands.

Figure 2 compares the tunneling $IV$ curves in the density balanced 5/2-5/2 and 7/2-7/2 configurations with the imbalanced 5/2-7/2 set-up.  The data in Fig. 2(a) were obtained at $T=50$ mK and \bperp\ = 0.84 T, with \bpar = 0.  Using the front and back electrostatic gates, the 2DES densities were adjusted to allow examination of these two different filling factors at the same magnetic field.  As the figure shows, under these conditions the $IV$ curves for the two filling factors are nearly identical.

In Fig. 2(b) the layer densities were adjusted to produce $\nu = 5/2$ in one layer and $\nu = 7/2$ in the other \cite{inverse}.  The tunnel junction is here configured such that positive interlayer voltage (`forward bias') raises the chemical potential of the 7/2 layer above that of the 5/2 layer while negative interlayer voltage (`reverse bias') does the opposite.  Two traces are shown in the figure, the blue one obtained at $T=50$ mK, the red at $T=200$ mK.  At $T=50$ mK the tunnel current over the range $|V| \lesssim 1$ mV is massively asymmetric; a large peak, centered at $V\approx 0.5$ mV, is seen at forward bias while in reverse bias only a very weak (negative) peak in the current is observed at $V \approx -0.25$ mV.  The tunnel junction at this low temperature is a robust quantum Hall spin diode.

The $T=50$ mK tunneling data in Fig. 2(b) imply that the spin polarization of the electrons (holes) in the $N=1$ LL is virtually total at $\nu=5/2$ ($\nu = 7/2$).  The tunnel current at the weak minimum near $V \approx -0.25$ mV is less than 1\% of that seen at the forward bias peak near $V\approx 0.5$ mV.  Moreover, since weak spin-flip tunneling processes (due, for example, to nuclear spin fluctuations) cannot be entirely ruled out, the electron spin polarization might be even more complete than these data suggest.  This finding, which is in agreement with recent Knight shift measurements \cite{tiemann12} of the spin polarization at $\nu = 5/2$, contrasts markedly with the situation at $\nu = 1/2$ in the $N=0$ LL where partial spin polarization is known to persist to magnetic fields considerably higher than those studied here \cite{kukushkin99,dementyev99,melinte00,dujovne05,tracy07,li09}. 

The peak in the tunnel currrent in forward bias occurs at about $V=0.5$ mV in the 5/2-7/2 data, but at about $V=0.25$ mV in the 5/2-5/2 and 7/2-7/2 cases.   This difference is expected since in the 5/2-7/2 case, alignment of same-spin sublevels in the two layers requires an additional voltage bias equal to the spin-flip energy $E_{sf}$.  Owing to exchange interactions, $E_{sf}$ typically greatly exceeds the ordinary Zeeman energy $E_Z$.   For the data in Fig. 2, $E_{sf}\approx 0.25$ meV, whereas $E_Z=0.021$ meV.

\begin{figure}
\includegraphics[width=3.25in]{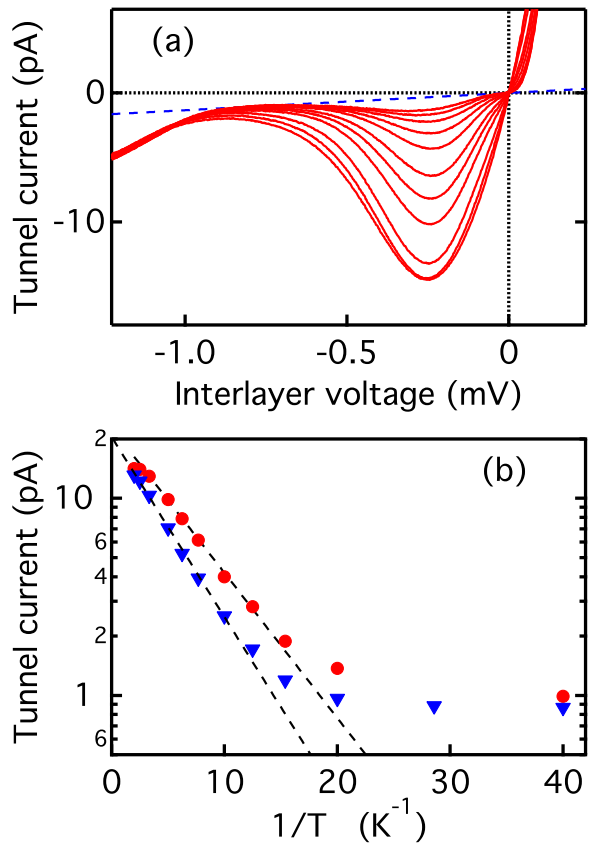}
\caption{\label{}(color online) (a) Temperature dependence of the reverse bias peak in 5/2-7/2 tunneling at \bperp\ = 0.84 T. Top to bottom: $T$ = 25, 50, 65, 80, 100, 130, 160, 200, 300, 400 and 500 mK.  Dashed line is estimated background.  (b) Arrhenius plots of reverse bias peak amplitude.  Red circles, \bperp\ =0.84 T; blue triangles: \bperp\ = 1.19 T.  Dashed lines suggest activation energies of 0.17 K and 0.21 K, respectively.}
\end{figure}
We turn finally to the temperature dependence of the tunneling spectra.  For the balanced 5/2-5/2 and 7/2-7/2 cases, and the 5/2-7/2 set-up in forward bias, little temperature dependence is observed for $T \lesssim 0.5$ K ($k_BT \lesssim 0.04$ meV).  This is not surprising given the broad resonances in these situations.  However, this contrasts sharply with the strong temperature dependence of the 5/2-7/2 spectrum in $reverse$ bias.  There, as Figs. 2(b) and 3(a) demonstrate, the (negative) peak in the tunnel current at $V \approx -0.25$ mV grows dramatically with temperature.  

Figure 3(b) shows the amplitude of the 5/2-7/2 reverse bias peak (with a small background subtracted) in an Arrhenius plot.  Two data sets are shown, one at $B_{\perp}=0.84$ T and the other at $B_{\perp}=1.19$ T.  At the lowest temperatures, both data sets saturate.  The saturation value may indicate a tiny, but genuine, lack of total spin polarization.  However, unknown spin-flip tunneling processes or a small amount of electron heating might also be responsible.  Above about $T=100$ mK, the peak amplitude is roughly thermally activated, with an activation energy of $E_A\approx 0.17$ K (0.21 K) for the $B_{\perp}$=0.84 T (1.19 T) data.  These activation energies are comparable to the bare Zeeman energy at the same magnetic fields ($E_Z=0.24$ K and 0.35 K, respectively).  Our data therefore suggest that thermal excitation of long-wavelength spin waves (whose energy $\epsilon_{sw} \rightarrow E_Z$ as $\lambda \rightarrow \infty$) is responsible for the temperature dependence of the reverse bias tunneling peak in the 5/2-7/2 spin diode configuration.  Both spin wave-assisted tunneling and, as has been suggested \cite{kasner96}, thermal fluctuations in the orientation of the spin polarization of the 2DES, provide plausible explanations for the enhanced reverse-bias tunneling currents our data have revealed.  Owing to these fluctuations, the spectral function of the 2DES no longer cleanly separates into spin-up and spin-down components.  Similar arguments have been used previously to understand the thermal behavior of magnetic tunnel junctions comprising three-dimensional itinerant ferromagnetic electrodes \cite{macdonald98}.

In conclusion, we have fabricated magnetic tunnel junctions from density imbalanced double layer 2D electron systems subjected to strong magnetic fields.  These junctions can behave as almost ideal spin diodes at low temperature, indicating that the spin polarization of the valence Landau level in each layer is nearly complete. Thermally excited spin waves appear to be responsible for degradation of diode behavior at elevated temperatures.  It should be possible engineer tunnel junctions in which one fully spin polarized layer is used to analyze an unknown spin configuration in the opposite layer.

It is a pleasure to acknowledge helpful discussions with P.A. Lee, A.H. MacDonald, R. Refael, S. Das Sarma, and A. Stern.
This work was supported in part by the Institute for Quantum Information and Matter, an NSF Physics Frontiers Center with support of the Gordon and Betty Moore Foundation through Grant No. GBMF1250.  The work at Princeton University was funded by the Gordon and Betty Moore Foundation through Grant GBMF 4420, and by the National Science Foundation MRSEC Grant 1420541.


\begin{references}
\bibitem{tedrow71} P. M. Tedrow and R. Meservey, Phys. Rev. Lett. {\bf 26}, 192 (1971).
\bibitem{meservey94} R. Meservey and P. M. Tedrow, Phys. Rep. {\bf 238}, 173 (1994). 
\bibitem{tedrow70} R. Meservey, P. M. Tedrow, and Peter Fulde, Phys. Rev. Lett. {\bf 25}, 1270 (1970).
\bibitem{singular} In this approximation the LLs are infinitely sharp.
\bibitem{eisenstein92} J. P. Eisenstein, L. N. Pfeiffer, and K. W. West, Phys. Rev. Lett. {\bf 68}, 3904 (1992).
\bibitem{ashoori90} R. C. Ashoori, J. A. Lebens, N. P. Bigelow, and R. H. Silsbee, Phys. Rev. Lett. {\bf 64}, 681 (1990).
\bibitem{hatsugai93} Y. Hatsugai, P. -A. Bares, and X. G. Wen, Phys. Rev. Lett. {\bf71}, 424 (1993).
\bibitem{he93} Song He, P. M. Platzman, and B. I. Halperin, Phys. Rev. Lett. {\bf71}, 777 (1993).
\bibitem{johannson93} Peter Johannson and Jari M. Kinaret, Phys. Rev. Lett. {\bf71}, 1435 (1993).
\bibitem{efros93} A. L. Efros and F. G. Pikus, Phys. Rev. B {\bf48}, 14694 (1993).
\bibitem{varma94} C. M. Varma, A. I. Larkin, and E. Abrahams, Phys.Rev. B {\bf 49}, 13999 (1994).
\bibitem{haussmann96} Rudolf Haussmann, Hiroyuki Mori, and A. H. MacDonald, Phys. Rev. Lett. {\bf 76}, 979 (1996).
\bibitem{levitov97} L. S. Levitov and A. V. Shytov, JETP Lett. {\bf66}, 214 (1997).
\bibitem{halperin83} This was first suggested by B. I. Halperin in Helv. Phys. Acta {\bf 57}, 75 (1983).
\bibitem{perspectives} For early examples, see {\it Perspectives on Quantum Hall Effects}, edited by S. Das Sarma and A. Pinczuk (Wiley, New York, 1997).
\bibitem{energies} The Zeeman energy is $E_Z=g\mu_B B$, with $g=-0.44$ the GaAs conduction band $g$-factor and $\mu_B$ the Bohr magneton.  The Coulomb energy is $E_C=e^2/\epsilon \ell$, with $\epsilon=13 \epsilon_0$ the GaAs dielectric constant and $\ell=(\hbar/eB_{\perp})^{1/2}$ the magnetic length.  
\bibitem{kukushkin99} I. V. Kukushkin, K. von Klitzing, and K. Eberl, Phys. Rev. Lett. {\bf 82}, 3665 (1999).
\bibitem{dementyev99} A. E. Dementyev, N. N. Kuzma, P. Khandelwal, S. E. Barrett, L. N. Pfeiffer, and K. W. West, Phys. Rev. Lett. {\bf 83}, 5074 (1999).
\bibitem{melinte00} S. Melinte, N. Freytag, M. Horvatic, C. Berthier, L. P. Levy, V. Bayot, and M. Shayegan, Phys. Rev. Lett. {\bf 84}, 354 (2000).
\bibitem{dujovne05} Irene Dujovne, A. Pinczuk, Moonsoo Kang, B. S. Dennis, L. N. Pfeiffer, and K. W. West, Phys. Rev. Lett. {\bf 95}, 056808 (2005).
\bibitem{tracy07} L.A. Tracy, J.P. Eisenstein. L.N. Pfeiffer, and K.W. West, Phys. Rev. Lett. {\bf 98}, 086801 (2007).
\bibitem{li09} Y. Q. Li, V. Umansky, K. von Klitzing, and J. H. Smet, Phys. Rev. Lett. {\bf 102}, 046803 (2009) and Phys. Rev. B {\bf 86}, 115421 (2012).
\bibitem{eisenstein90} J. P. Eisenstein, L. N. Pfeiffer, and K. W. West, Appl. Phys. Lett. {\bf 57}, 2324 (1990).
\bibitem{eisenstein91} J. P. Eisenstein, L. N. Pfeiffer, and K. W. West, Appl. Phys. Lett. {\bf 58}, 1497 (1991).
\bibitem{asymmetry} The deviation of the $IV$ curves in Fig. 1 from perfect antisymmetry in $V$ is not understood.  While layer density imbalance in the bilayer 2DES might be responsible, the imbalance required is inconsistent with the precision balancing of the sample provided by gate-tuning the tunnel resonance at zero magnetic field and the lack of any detectable beating in the magneto-oscillations of the tunnel conductance at low magnetic field.
\bibitem{forbidden} Provided that the magnetic field is perpendicular to the 2D planes.
\bibitem{disorder} Both disorder and electron-electron interactions can enable inter-LL tunneling.
 \bibitem{hu92} J. Hu and A. H. MacDonald, Phys. Rev. B {\bf 46}, 12554 (1992).
\bibitem{eisenstein16} In the $N=0$ Landau level $I(B_{||})=I(0) e^{-q^2\ell^2/2}$.  See J. P. Eisenstein, T. Khaire, D. Nandi, A. D. K. Finck, L. N. Pfeiffer, and K. W. West, Phys. Rev. B {\bf 94}, 125409 (2016) for a demonstration.
\bibitem{maxdef} The maximum tunnel current is taken as the average of the peak heights at positive and negative interlayer voltage $V$.
\bibitem{inverse} For the data in Fig. 3(b), the upper 2DES is at $\nu=5/2$ and the lower at $\nu=7/2$.  Essentially identical data were obtained when this situation was reversed.
\bibitem{tiemann12} Lars Tiemann, Gerardo Gamez, Norio Kumada, Koji Muraki, Science {\bf 335}, 828 (2012).
\bibitem{kasner96} Marcus Kasner and A. H. MacDonald, Phys. Rev. Lett. {\bf 76}, 3204 (1996) and private communication.
\bibitem{macdonald98} A. H. MacDonald, T. Jungwirth, and M. Kasner, Phys. Rev. Lett. {\bf 81}, 705 (1998).


\end{references}
\end{document}